\documentclass[aps,prx,twocolumn,amsmath,amssymb,showpacs]{revtex4-1}

\usepackage{siunitx}
\usepackage[ansinew]{inputenc}
\usepackage{graphicx}
\usepackage{textcomp}
\usepackage{amsmath}
\usepackage{amssymb}
\usepackage{xcolor}
\usepackage{braket}
\newcommand{\dd}{\mathrm{d}}

\begin{document}

\title{Laser stabilization to a cryogenic fiber ring resonator}
\author{Benjamin Merkel}
\author{Daniel Repp}
\author{Andreas Reiserer}
\email{andreas.reiserer@mpq.mpg.de}

\affiliation{Quantum Networks Group, Max-Planck-Institut f\"ur Quantenoptik, Hans-Kopfermann-Strasse 1, D-85748 Garching, Germany}
\affiliation{Munich Center for Quantum Science and Technology (MCQST), Ludwig-Maximilians-Universit\"at
M\"unchen, Fakult\"at f\"ur Physik, Schellingstr. 4, D-80799 M\"unchen, Germany}

\begin{abstract}
The frequency stability of lasers is limited by thermal noise in state-of-the-art frequency references. Further improvement requires operation at cryogenic temperature. In this context, we investigate a fiber-based ring resonator. Our system exhibits a first-order temperature-insensitive point around $3.55\,\si{\kelvin}$, much lower than that of crystalline silicon. The observed low sensitivity with respect to vibrations  ($<5\cdot{10^{-11}}\,\si{m^{-1} s^{2}}$), temperature ($-22(1)\cdot{10^{-9}}\,\si{\kelvin^{-2}}$) and pressure changes ($4.2(2)\cdot{10^{-11}}\,\si{\milli\bar^{-2}}$) makes our approach promising for future precision experiments.
\end{abstract}

\maketitle

The construction of lasers with ultra-high frequency stability is a key enabling technology for optical atomic clocks \cite{ludlow_optical_2015} and a large variety of precision measurements \cite{krohn_fiber_2014, ghelfi_fully_2014, derevianko_hunting_2014, hogan_atom-interferometric_2016, canuel_exploring_2018, karr_progress_2019}. In spite of recent alternative approaches \cite{norcia_frequency_2018}, the best results have been obtained with lasers locked to ultra-stable external frequency references. Well-studied examples for such system include spectral holes in rare-earth doped crystals \cite{thorpe_frequency_2011,cook_laser-frequency_2015}, fiber-optical delay lines \cite{kefelian_ultralow-frequency-noise_2009,dong_subhertz_2015, dong_observation_2016}, as well as whispering-gallery-mode \cite{lim_chasing_2017} and Fabry-Perot resonators \cite{salomon_laser_1988, webster_thermal-noise-limited_2008, kessler_sub-40-mhz-linewidth_2012}. The latter have demonstrated an impressive performance down to a relative frequency stability of $4\cdot10^{-17}$ \cite{matei_1.5_2017}. In that experiment, the detrimental effect of temperature fluctuations has been minimized by operating a crystalline silicon resonator at a zero-crossing of its thermal expansion coefficient around $124\,\si{\kelvin}$. The achieved frequency stability has then been limited by thermal noise of the mirror coatings \cite{numata_thermal-noise_2004}. While this noise can be reduced by cooling to even lower temperature \cite{zhang_ultrastable_2017, robinson_crystalline_2019}, this comes at the prize of a larger sensitivity to temperature drifts, as the system is then operated below the zero-crossing of the thermal expansion coefficient of silicon. 

Therefore, in this work we explore a cryogenic fiber-based resonator as an alternative design for a frequency-stable reference. This has two advantages: First, such resonator is easier to implement as it only requires off-the-shelf components. Second, in this work we show that our fiber resonator exhibits a temperature-insensitive point around 3.55\,\si{\kelvin}, 35-fold lower than that of crystalline silicon studied previously \cite{zhang_ultrastable_2017}.

Fiber delay lines have been investigated at room temperature, mainly because of their lower cost and complexity \cite{kefelian_ultralow-frequency-noise_2009} as compared to Fabry-Perot resonators. Recent experiments have demonstrated sub-Hz short-term stability \cite{dong_subhertz_2015}. However, both temperature fluctuations and thermal noise have been limiting the performance \cite{dong_observation_2016}. Here we show that both of these limitations can be alleviated when operating at cryogenic temperature.

Our experiment is intended as a proof-of-concept. We thus do not target or achieve the ultra-high precision of other cryogenic experiments with Fabry-Perot resonators \cite{zhang_ultrastable_2017} or rare-earth doped crystals \cite{thorpe_frequency_2011, cook_laser-frequency_2015}. Still, we perform a detailed characterization of our device with respect to vibrations as well as temperature and pressure instability. The observed low sensitivity and in particular the existence of a temperature-insensitive point make our approach promising for future precision experiments.

Our experiment uses a fiber ring resonator which exhibits a ladder of equidistant resonances with free spectral range $f_\text{FSR}=c/(nL)$. Here, $n$ is the refractive index, $L$ the length of the fiber, and $c$ the speed of light. In most Fabry-Perot frequency references investigated to date \cite{matei_second_2016, webster_thermal-noise-limited_2008} the cavity field is in vacuum with constant $n=1$. Thus, a temperature-insensitive point is observed when the thermal expansion coefficient $\alpha=L^{-1}(\partial L /\partial T)$ exhibits a zero-crossing. In contrast, in our experiment the light is guided in a silica fiber, whose refractive index changes with temperature $T$, as described by the normalized thermo-optic coefficient $\beta_\text{TO}=n^{-1}(\partial n/\partial T)$. 
In our setting, both the thermal expansion \cite{white_thermal_1975} and the thermo-optic coefficient \cite{arcizet_cryogenic_2009} change with temperature. The combined sensitivity $\alpha+\beta_\text{TO}$ of fused silica exhibits a zero-crossing around $13\,\si{K}$ \cite{arcizet_cryogenic_2009}, making it a promising material for cryogenic frequency references.

Instead of pure glass, we use a commercial fiber, which gives an additional contribution to the total temperature sensitivity. The fiber cladding will exert a radial pressure on the core if their thermal expansion coefficients are not the same, which changes both the refractive index and fiber length. The effect on phase stability has been studied recently down to temperatures $100\,\si{\kelvin}$ \cite{zhu_thermal_2020}, finding that acrylate coatings transition to a stiff, glass-like state with thermal expansion converging to zero at $0\,\si{\kelvin}$. In our modeling, we include the effect of thermal strain as coefficient $\beta_\text{TS}$, getting the total temperature sensitivity:

\begin{equation} \label{eq:df_dT_over_f}
    \frac{1}{f} \frac{\dd f}{\dd T} = -(\alpha + \beta_\text{TO} + \beta_\text{TS}).
\end{equation}

The existence, temperature and curvature of a temperature-insensitive point, which corresponds to a zero-crossing of eq.~\ref{eq:df_dT_over_f}, will thus depend on the used fiber. In particular the material and diameter of the core, cladding, and acrylate coating will determine the thermal strain coefficient. While we use a standard commercial product for our initial experiments, this gives access to a large parameter space for future optimization.

In our experiment, we fabricate a fiber ring resonator by splicing the ends of a 95:5 fused fiber beam-splitter to a $\sim 120\,\si{m}$ long fiber. The latter is coiled to an aluminum cylinder of $4\,\si{\centi\metre}$ outer diameter that fits into the sample space of our closed-cycle cryostat (AttoDry 2100). To avoid bend-induced loss that would limit the finesse, both the beam-splitter and fiber are made from Corning$^\text{\textregistered}$ ClearCurve$^\text{\textregistered}$ bend-insensitive fiber. We do not expect that the properties of the fused coupler (Evanescent Optics Inc., type 954) significantly influence our measurements.

\begin{figure}
\includegraphics[width=1\columnwidth]{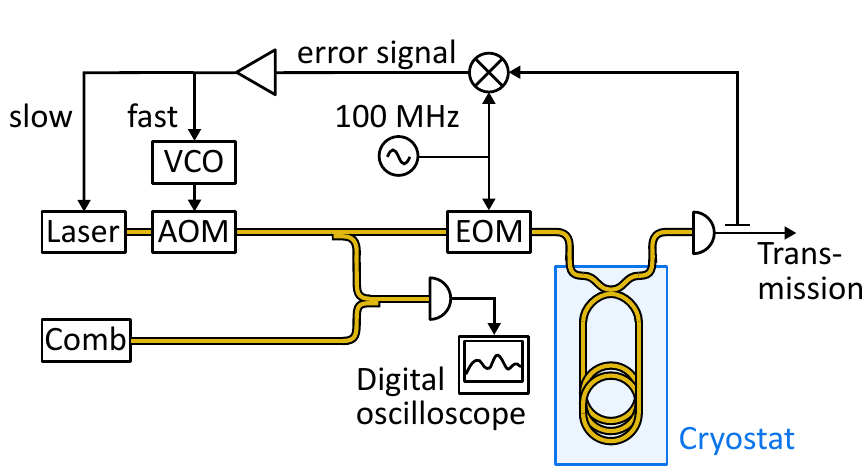}
\caption{\textbf{Experimental setup.}\label{fig:setup}
A cryogenic fiber ring resonator serves as a frequency reference for stabilizing a laser at $1535\,\si{nm}$. To this end, the transmission is measured by a fast photodiode. An error signal is generated using the Pound-Drever-Hall technique with the help of a radio-frequency source at 100 MHz, an electro-optical modulator (EOM) and a mixer. The feedback signal is applied to a voltage-controlled oscillator (VCO) that drives an acousto-optical modulator (AOM). In addition, slow drifts of the laser frequency are compensated via its tuning port. To characterize the frequency stability, the beating signal of the laser light with an ultra-stable frequency comb is recorded.
}

\end{figure}

We first determine the resonator properties at cryogenic temperature by measuring its transmission at $1535\,\si{nm}$ through the two open ports of the beam-splitter, c.f. Fig.~\ref{fig:setup}. After adjusting the input polarization to match one of the resonator eigenmodes, we observe a free spectral range of $1.71(4)\,\si{MHz}$ and a FWHM linewidth of $87(7)\,\si{kHz}$. This corresponds to a finesse of 20(1) and a round-trip transmission of about $70\,\%$. This value is limited by the splice-, bend- and absorption loss of the fiber and the excess loss of the beam splitter, and might be further increased in future devices. The sample is thermalized to its cryogenic environment via helium exchange gas, whose pressure is monitored with a Pirani pressure gauge at ambient temperature. To characterize the sensitivity of the device to perturbations, we attach a cryogenic vibration sensor and a resistive thermometer in close proximity to the ring resonator.

\begin{figure}
\includegraphics{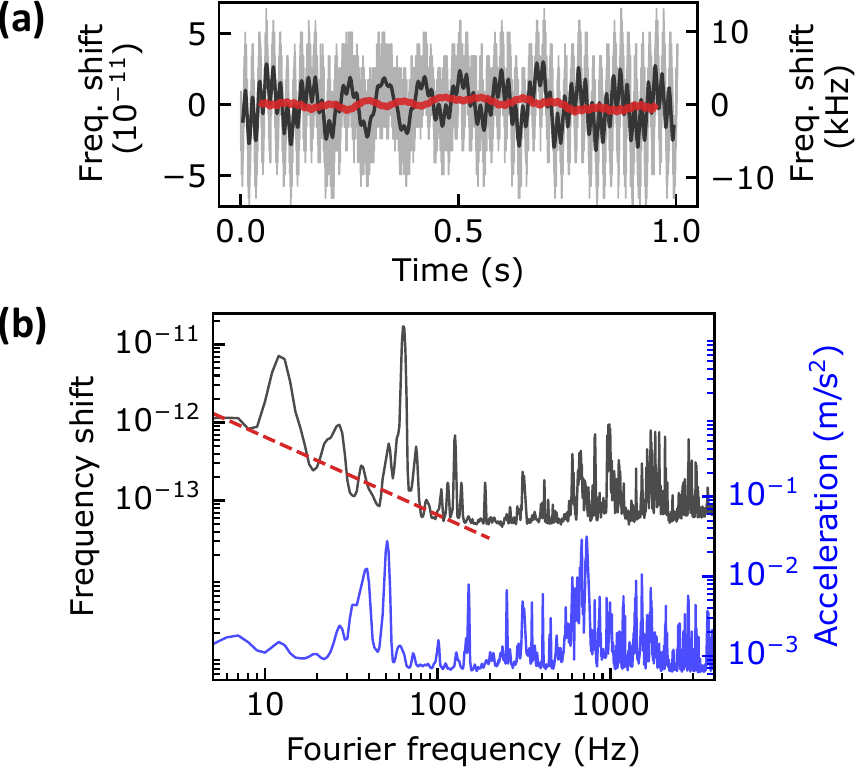}
\caption{\textbf{Short-term stability.}
(a) Frequency shift of the resonance as a function of time (grey), with running averages over $10\,\si{ms}$ (black) and $100\,\si{ms}$ (red). 
(b) Transmission spectrum (black) of the fiber ring resonator obtained from the data in (a), compared to the sample stage vibrations (blue) measured by a cryogenic acceleration sensor. The red dashed line indicates the calculated noise floor caused by measured temperature drifts of about $1\,\si{\milli\kelvin\per\second}$. 
}
\label{fig:shortterm}
\end{figure}

We now study the stability of the resonator  against external perturbations. To investigate its short-term stability, we lock the laser (Koheras BASIK X15) to a frequency comb (Menlo Systems FC1500-250-ULN) that in turn is referenced to a resonator (Menlo Systems ORS1500) with sub-Hz stability. We then tune the laser to the side of the fiber resonator transmission dip. Fluctuations of the resonator frequency will lead to a fluctuation of the transmitted power, which we transfer to frequency deviations using the independently measured spectral response. The resulting time trace is shown in Fig.~\ref{fig:shortterm}(a). A fast-Fourier-transform gives the spectral properties of the frequency shift (b), which exhibits a number of peaks at different frequencies (black). The reason for the peaked structure is acoustic resonances that are excited by the broadband noise of our pulse tube cryocooler. The position and width of the resonances shows large similarities with the vibration spectrum measured by the attached piezo-electric sensor (blue). Deviations in the amplitude of the peaks can be explained by the sensor being only sensitive to vibrations along the axis parallel to the coil center, while the resonator will be sensitive to vibrations along all axes. By comparing the peaks in the spectra of sensor and resonator between $0.3$ and $4\,\si{\kilo\hertz}$, we can estimate that the vibration sensitivity of our resonator is about $<5\cdot{10^{-11}}\,\si{m^{-1}\,s^{2}}$ in that frequency range.

\begin{figure}[ht]
\centering\includegraphics[width=1\columnwidth]{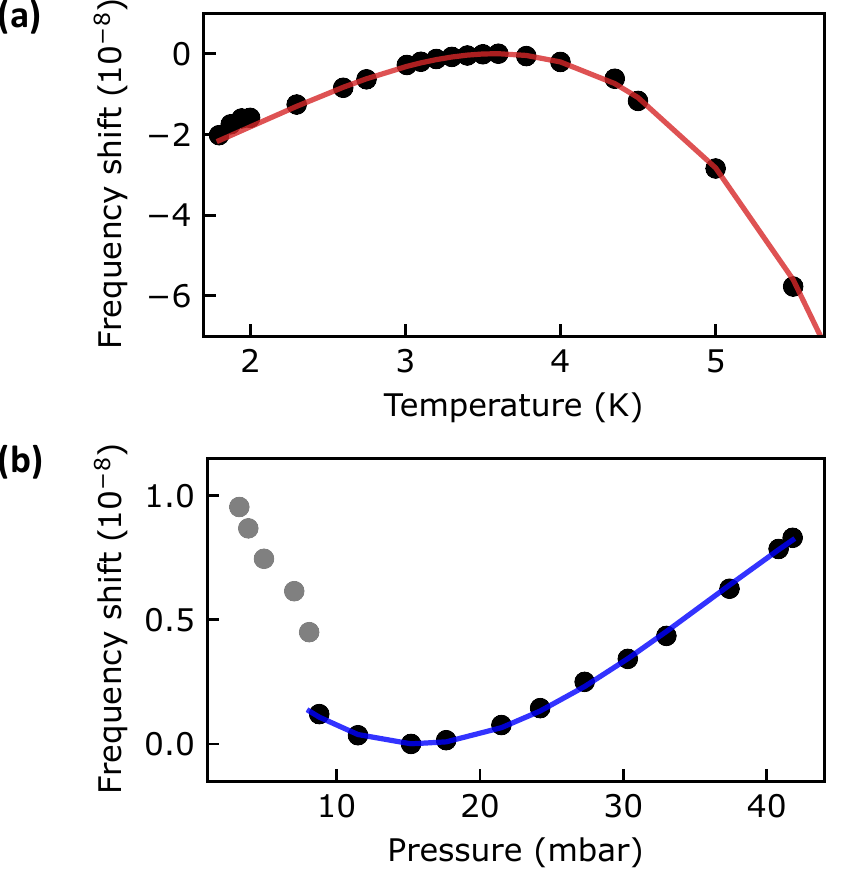}
\caption{\textbf{Temperature and pressure sensitivity.} 
\textbf{Main graph:} The frequency shift of the fiber ring resonances with temperature shows a turning point around $3.55\,\si{K}$, shown for an exchange gas pressure of $18\,\si{mbar}$. The measured data (black dots) are fit by a third-order polynomial (red line) to extract a curvature of $-22(1)\cdot{10^{-9}}\,\si{K^{-2}}$ at the turning point. 
\textbf{Inset:} Shift of the resonance frequency with pressure at a temperature of $3.55\,\si{K}$. At low pressure (grey dots) the resonator thermalization may be impaired. A fit to a third-order polynomial (blue line) gives a pressure-insensitive point around 15 mbar, with a remaining pressure dependence of $4.2(2)\cdot{10^{-11}}\,\si{\milli\bar^{-2}}$.}
\label{fig:TP_dependence}
\end{figure}

As a next step, we investigate the sensitivity  with respect to temperature and pressure changes by locking the laser to the resonator using the Pound-Drever-Hall technique \cite{black_introduction_2001}, with an input power of $\sim 0.1\,\si{\milli\watt}$. We use a modulation frequency that is much larger than the free-spectral range and tuned such that the sidebands are halfway between higher-order resonances and thus not phase-shifted upon transmission. We then record the frequency difference between the laser and frequency comb after changing the sample space temperature and waiting a few seconds until full thermalization.

The obtained temperature dependence (with a constant amount of helium in the sample space) is shown in Fig.~\ref{fig:TP_dependence}(a). We observe a first-order temperature-insensitive point around $3.55\,\si{\kelvin}$, (Fig.~\ref{fig:TP_dependence}(a)). The data is well-fit by a third-order polynomial (red curve), from which we extract the curvature at the turning point, $-22(1)\cdot{10^{-9}}\,\si{K^{-2}}$. We repeat this procedure for different amounts of helium gas in the sample space and thus different pressures. We find that the temperature of the turning point and its curvature do not change significantly (not shown). 

To directly investigate the pressure sensitivity, we evacuate the sample space and then repeatedly add small amounts of helium while the temperature of the sample space is kept at the insensitive point, $\sim 3.55\,\si{\kelvin}$. 
The observed frequency shift (Fig.~\ref{fig:TP_dependence}(b)) corresponds to a sensitivity of $\lesssim 5\cdot{10^{-10}}\,\si{mbar^{-1}}$ over a large pressure range, with a first-order insensitive point around $15\,\si{\milli\bar}$. Measurements at lower pressure (grey) are less reliable as the thermalization of the fiber resonator with the surroundings is impaired when the pressure is too low. Similarly, we cannot exclude that changing temperature gradients contribute to the measured pressure dependence. Thus, the above value should be considered as an upper bound. Still, we note that the observed low pressure sensitivity justifies that our temperature scans have been performed at constant helium filling level instead of constant gas pressure. The reason is that in the investigated regime, the pressure changes at $\lesssim 2.5\,\si{mbar/K}$. Thus, its impact on the temperature sensitivity is only $\lesssim 1.25\cdot{10^{-9}}\,\si{K^{-1}}$, i.e. small compared to the observed temperature dependence. For the same reason, compensating temperature changes by connecting the sample space to a room-temperature helium reservoir, as pioneered in \cite{cook_laser-frequency_2015}, does not seem promising in our fiber-based ring resonator unless the temperature sensitivity can be further reduced by materials engineering.

\begin{figure}
\centering\includegraphics{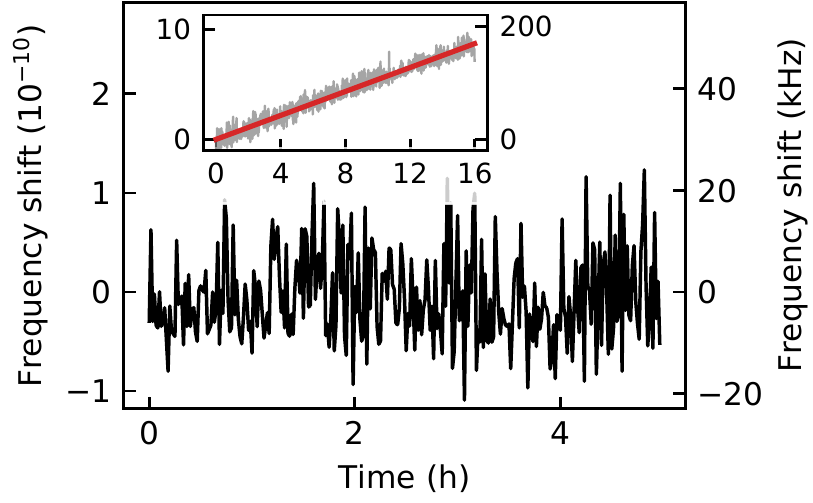}
\caption{\textbf{Long-term stability.}
Shift of the laser frequency when locked to the fiber ring resonator (with $60\,\si{s}$ averaging intervals to eliminate the influence of vibrations). The temperature is kept within $100\,\si{mK}$ around the temperature-insensitive point of $3.55\,\si{K}$. The inset shows the raw data, while the main graph has been obtained by subtracting a linear fit.}
\label{fig:longterm}
\end{figure}

After characterizing the sensitivity to external perturbations, we measured the stability over a period of sixteen hours, see Fig.~\ref{fig:longterm}. The inset shows the raw data with a slow linear drift by about $5\cdot{10^{-11}}\,\si{h^{-1}}$, similar to the isothermal creep reported for other amorphous materials such as ultra-low expansion glass commonly used in reference cavities \cite{webster_thermal-noise-limited_2008}. As it may originate from the thermal stress exerted by the fiber cladding, different fiber types may show a different linear drift. Still, after subtracting a linear fit, our resonator exhibits a long-term stability around $20\,\si{kHz}$, or $\cdot{10^{-10}}$, limited by the moderate temperature stability of $\pm 100\,\si{mK}$ obtained in our cryostat.


Thus, our implementation does not achieve a  stability that exceeds previous experiments, neither those with fiber interferometers at room temperature ($5\cdot 10^{-15}$) \cite{dong_subhertz_2015}, nor that of cryogenic silicon resonators ($4\cdot10^{-17}$) \cite{matei_1.5_2017}. The reason is that our setup does not include any thermal shields or vibration-damping enclosures. Instead, it constitutes a proof-of-concept experiment to determine the sensitivity to pressure, temperature and vibrations, which we discuss in the following.

The extracted vibration sensitivity of our fiber coil, $<5\cdot{10^{-11}}\,\si{(m^{-1}\,s^{2})}$, is only about tenfold larger than that of specially optimized, unitary aspect ratio ultra-low expansion glass cavities at room temperature \cite{webster_thermal-noise-limited_2008, leibrandt_spherical_2011} and that of crystalline silicon resonators \cite{matei_1.5_2017}. Operating our resonator in a closed-cycle cryocooler with decoupled vibrations down to a level of $\lesssim 10^{-3}\,\si{m/s^2}$ \cite{zhang_ultrastable_2017}, we expect that our fiber ring resonator would perform around $10^{-14}$ short-term stability. Even better short-term stability can be achieved with improvements of our resonator design. First, additional vibration damping material may be inserted between the aluminum cylinder and the fiber coil. Second, increasing the fiber length while reducing the finesse may be advantageous. Third, an optimized geometric arrangement of the fiber may reduce the vibration sensitivity, with a fifty-fold lower value demonstrated at room temperature \cite{huang_optical_2019}. Finally, our setup does not require coupling of a free-space optical beam into a micro-sized resonator mode. Thus, it should be possible to build an effective cryogenic vibration isolation stage. As the thermalization is done by exchange gas, one could e.g. simply use a large mass on a soft spring to hold the fiber resonator while damping out all high-frequency vibrations. Alternatively, active damping and magnetic levitation could be implemented in cryostats with larger sample space, offering the potential for unprecedented short-term stability.

In this context, also the upper bound of the pressure sensitivity, $4.2(2)\cdot{10^{-11}}\,\si{\milli\bar^{-2}}$, is important. It is much smaller than the linear change of $\sim 3\cdot{10^{-7}}\,\si{\milli\bar^{-2}}$ observed with Fabry-Perot cavities at room temperature and atmospheric pressure \cite{egan_performance_2015}. Getting the stability to the $10^{-16}$ level would require pressure stabilization to $10^{-4}$, which should be straightforward in a closed cryogenic volume. Alternatively, placing the system in cryogenic vacuum with typical pressures $\ll 10^{-9}\,\si{\milli\bar}$ will eliminate the influence of pressure and its fluctuations.

Next, we compare the observed temperature sensitivity of $22(1)\cdot{10^{-9}}\,\si{K^{-2}}$ to other experiments. It is about an order of magnitude worse than ultra-low expansion glass cavities at room temperature, $1.5\cdot{10^{-9}}\,\si{K^{-2}}$ \cite{webster_thermal-noise-limited_2008}, but close to that of Fabry-Perot cavities made from crystalline silicon and operated at their temperature-insensitive point,  $17\cdot{10^{-9}}\,\si{K^{-2}}$ at $124\,\si{K}$ \cite{kessler_sub-40-mhz-linewidth_2012}. 

Therefore, to estimate the potential of our cryogenic fiber ring resonator, we make an explicit comparison with the most stable sub-10\,K resonator to date: a silicon cavity operated around $4\,\si{K}$ \cite{zhang_ultrastable_2017, robinson_crystalline_2019}. To achieve the same linear sensitivity ($0.02\,\si{ppb/K}$), we would need to stabilize the fiber within $1\,\si{mK}$ proximity to the turning point, which is directly feasible in most commercial cryocoolers. With additional passive and active heat shields, temperature fluctuations below $10\,\si{\micro K}$ have been demonstrated \cite{zhang_ultrastable_2017}. For our resonator, this would lead to an expected stability below $2\cdot 10^{-18}$. Optimization of the fiber coating thickness and material may allow for even lower temperature sensitivity at the turning point.

Still, the question whether such setup would allow for unprecedented frequency stability requires a careful analysis of thermal noise in the fiber. There are two main mechanisms predicted from theory \cite{wanser_fundamental_1992, duan_intrinsic_2010} and confirmed experimentally \cite{dong_subhertz_2015, dong_observation_2016}: At high frequencies, thermodynamic noise (thermoelastic and thermorefractive) \cite{wanser_fundamental_1992} dominates the spectrum in room temperature experiments. This contribution scales with $T^2\frac{\dd f}{\dd T}$. When operating at cryogenic temperature, and in particular at the temperature insensitive point, it should therefore be negligible. The second cause of thermal noise can be derived from the fluctuation-dissipation theorem, where the spectral density of spontaneous fiber length fluctuations reads \cite{duan_intrinsic_2010}:
\begin{equation}
    S_l(f)=\frac{2 k T L \Phi_0}{3 \pi E_0 A f}.
\end{equation}
Here, $k$ is Boltzmann's constant, $L$ the fiber length and $A$ its cross section, $f$ the frequency, $E_0$ the value of Young's modulus without loss and $\Phi_0$ its loss angle. The $T$-scaling in the above formula suggests that cooling a fiber to cryogenic temperature may lead to an improved stability -- comparing to measurements at room temperature \cite{dong_subhertz_2015} even linewidths of a few mHz seem feasible. However, care has to be taken in this extrapolation, as the loss angle of the fiber assembly will also change when lowering the temperature. While that of the fiber will likely increase \cite{arcizet_cryogenic_2009}, that of the acrylate coating may be reduced when it transitions to a stiff, glass-like state \cite{zhu_thermal_2020}. Therefore, additional measurements are required to give a reliable estimation about the ultimately achievable stability of a cryogenic fiber-based setup.

In summary, we have characterized the sensitivity of a fiber-ring resonator to environmental fluctuations at cryogenic temperature. Our approach may find direct application in laboratories that operate cryogenic setups and have moderately high requirements on laser stability. As an example, our setup is intended for spectroscopy of rare-earth doped crystals with lifetime-limited spectral resolution \cite{merkel_coherent_2020}. In addition, the robustness and light-weight design of our resonator makes it promising for laser stabilization in space \cite{mcrae_frequency_2013}. Finally, when operated in an optimized closed-cycle cryostat \cite{zhang_ultrastable_2017}, our system may also be considered for laser stabilization to an unprecedented accuracy, depending on the yet unknown contribution of thermal noise at cryogenic temperature. If this contribution is too large, however, further reduction by two orders of magnitude seems feasible by operating our setup in a dilution refrigerator.

\section*{Acknowledgments}
This project received funding from the European Research Council (ERC) under the European Union's Horizon 2020 research and innovation programme (grant agreement No 757772), and from the Deutsche Forschungsgemeinschaft (DFG, German Research Foundation) under Germany's Excellence Strategy - EXC-2111 - 390814868. We acknowledge the technical contribution of Samarth Chawla during an early stage of the project, and discussions with Thomas Legero.

\bibliography{bibliography.bib}

\begin{thebibliography}{33}%
\makeatletter
\providecommand \@ifxundefined [1]{%
 \@ifx{#1\undefined}
}%
\providecommand \@ifnum [1]{%
 \ifnum #1\expandafter \@firstoftwo
 \else \expandafter \@secondoftwo
 \fi
}%
\providecommand \@ifx [1]{%
 \ifx #1\expandafter \@firstoftwo
 \else \expandafter \@secondoftwo
 \fi
}%
\providecommand \natexlab [1]{#1}%
\providecommand \enquote  [1]{``#1''}%
\providecommand \bibnamefont  [1]{#1}%
\providecommand \bibfnamefont [1]{#1}%
\providecommand \citenamefont [1]{#1}%
\providecommand \href@noop [0]{\@secondoftwo}%
\providecommand \href [0]{\begingroup \@sanitize@url \@href}%
\providecommand \@href[1]{\@@startlink{#1}\@@href}%
\providecommand \@@href[1]{\endgroup#1\@@endlink}%
\providecommand \@sanitize@url [0]{\catcode `\\12\catcode `\$12\catcode
  `\&12\catcode `\#12\catcode `\^12\catcode `\_12\catcode `\%12\relax}%
\providecommand \@@startlink[1]{}%
\providecommand \@@endlink[0]{}%
\providecommand \url  [0]{\begingroup\@sanitize@url \@url }%
\providecommand \@url [1]{\endgroup\@href {#1}{\urlprefix }}%
\providecommand \urlprefix  [0]{URL }%
\providecommand \Eprint [0]{\href }%
\providecommand \doibase [0]{http://dx.doi.org/}%
\providecommand \selectlanguage [0]{\@gobble}%
\providecommand \bibinfo  [0]{\@secondoftwo}%
\providecommand \bibfield  [0]{\@secondoftwo}%
\providecommand \translation [1]{[#1]}%
\providecommand \BibitemOpen [0]{}%
\providecommand \bibitemStop [0]{}%
\providecommand \bibitemNoStop [0]{.\EOS\space}%
\providecommand \EOS [0]{\spacefactor3000\relax}%
\providecommand \BibitemShut  [1]{\csname bibitem#1\endcsname}%
\let\auto@bib@innerbib\@empty
\bibitem [{\citenamefont {Ludlow}\ \emph {et~al.}(2015)\citenamefont {Ludlow},
  \citenamefont {Boyd}, \citenamefont {Ye}, \citenamefont {Peik},\ and\
  \citenamefont {Schmidt}}]{ludlow_optical_2015}%
  \BibitemOpen
  \bibfield  {author} {\bibinfo {author} {\bibfnamefont {A.~D.}\ \bibnamefont
  {Ludlow}}, \bibinfo {author} {\bibfnamefont {M.~M.}\ \bibnamefont {Boyd}},
  \bibinfo {author} {\bibfnamefont {J.}~\bibnamefont {Ye}}, \bibinfo {author}
  {\bibfnamefont {E.}~\bibnamefont {Peik}}, \ and\ \bibinfo {author}
  {\bibfnamefont {P.~O.}\ \bibnamefont {Schmidt}},\ }\href {\doibase
  10.1103/RevModPhys.87.637} {\bibfield  {journal} {\bibinfo  {journal} {Rev.
  Mod. Phys.}\ }\textbf {\bibinfo {volume} {87}},\ \bibinfo {pages} {637}
  (\bibinfo {year} {2015})}\BibitemShut {NoStop}%
\bibitem [{\citenamefont {Krohn}\ \emph {et~al.}(2014)\citenamefont {Krohn},
  \citenamefont {MacDougall},\ and\ \citenamefont {Mendez}}]{krohn_fiber_2014}%
  \BibitemOpen
  \bibfield  {author} {\bibinfo {author} {\bibfnamefont {D.~A.}\ \bibnamefont
  {Krohn}}, \bibinfo {author} {\bibfnamefont {T.~W.}\ \bibnamefont
  {MacDougall}}, \ and\ \bibinfo {author} {\bibfnamefont {A.}~\bibnamefont
  {Mendez}},\ }\href
  {http://ebooks.spiedigitallibrary.org/book.aspx?doi=10.1117/3.1002910} {\emph
  {\bibinfo {title} {Fiber {Optic} {Sensors}: {Fundamentals} and
  {Applications}}}}\ (\bibinfo  {publisher} {SPIE},\ \bibinfo {year}
  {2014})\BibitemShut {NoStop}%
\bibitem [{\citenamefont {Ghelfi}\ \emph {et~al.}(2014)\citenamefont {Ghelfi},
  \citenamefont {Laghezza}, \citenamefont {Scotti}, \citenamefont {Serafino},
  \citenamefont {Capria}, \citenamefont {Pinna}, \citenamefont {Onori},
  \citenamefont {Porzi}, \citenamefont {Scaffardi}, \citenamefont {Malacarne},
  \citenamefont {Vercesi}, \citenamefont {Lazzeri}, \citenamefont {Berizzi},\
  and\ \citenamefont {Bogoni}}]{ghelfi_fully_2014}%
  \BibitemOpen
  \bibfield  {author} {\bibinfo {author} {\bibfnamefont {P.}~\bibnamefont
  {Ghelfi}}, \bibinfo {author} {\bibfnamefont {F.}~\bibnamefont {Laghezza}},
  \bibinfo {author} {\bibfnamefont {F.}~\bibnamefont {Scotti}}, \bibinfo
  {author} {\bibfnamefont {G.}~\bibnamefont {Serafino}}, \bibinfo {author}
  {\bibfnamefont {A.}~\bibnamefont {Capria}}, \bibinfo {author} {\bibfnamefont
  {S.}~\bibnamefont {Pinna}}, \bibinfo {author} {\bibfnamefont
  {D.}~\bibnamefont {Onori}}, \bibinfo {author} {\bibfnamefont
  {C.}~\bibnamefont {Porzi}}, \bibinfo {author} {\bibfnamefont
  {M.}~\bibnamefont {Scaffardi}}, \bibinfo {author} {\bibfnamefont
  {A.}~\bibnamefont {Malacarne}}, \bibinfo {author} {\bibfnamefont
  {V.}~\bibnamefont {Vercesi}}, \bibinfo {author} {\bibfnamefont
  {E.}~\bibnamefont {Lazzeri}}, \bibinfo {author} {\bibfnamefont
  {F.}~\bibnamefont {Berizzi}}, \ and\ \bibinfo {author} {\bibfnamefont
  {A.}~\bibnamefont {Bogoni}},\ }\href {\doibase 10.1038/nature13078}
  {\bibfield  {journal} {\bibinfo  {journal} {Nature}\ }\textbf {\bibinfo
  {volume} {507}},\ \bibinfo {pages} {341} (\bibinfo {year}
  {2014})}\BibitemShut {NoStop}%
\bibitem [{\citenamefont {Derevianko}\ and\ \citenamefont
  {Pospelov}(2014)}]{derevianko_hunting_2014}%
  \BibitemOpen
  \bibfield  {author} {\bibinfo {author} {\bibfnamefont {A.}~\bibnamefont
  {Derevianko}}\ and\ \bibinfo {author} {\bibfnamefont {M.}~\bibnamefont
  {Pospelov}},\ }\href {\doibase 10.1038/nphys3137} {\bibfield  {journal}
  {\bibinfo  {journal} {Nat. Phys.}\ }\textbf {\bibinfo {volume} {10}},\
  \bibinfo {pages} {933} (\bibinfo {year} {2014})}\BibitemShut {NoStop}%
\bibitem [{\citenamefont {Hogan}\ and\ \citenamefont
  {Kasevich}(2016)}]{hogan_atom-interferometric_2016}%
  \BibitemOpen
  \bibfield  {author} {\bibinfo {author} {\bibfnamefont {J.~M.}\ \bibnamefont
  {Hogan}}\ and\ \bibinfo {author} {\bibfnamefont {M.~A.}\ \bibnamefont
  {Kasevich}},\ }\href {\doibase 10.1103/PhysRevA.94.033632} {\bibfield
  {journal} {\bibinfo  {journal} {Phys. Rev. A}\ }\textbf {\bibinfo {volume}
  {94}},\ \bibinfo {pages} {033632} (\bibinfo {year} {2016})}\BibitemShut
  {NoStop}%
\bibitem [{\citenamefont {Canuel}\ \emph {et~al.}(2018)\citenamefont {Canuel},
  \citenamefont {Bertoldi}, \citenamefont {Amand}, \citenamefont {Pozzo~di
  Borgo}, \citenamefont {Chantrait}, \citenamefont {Danquigny}, \citenamefont
  {Dovale~{\'A}lvarez}, \citenamefont {Fang}, \citenamefont {Freise},
  \citenamefont {Geiger}, \citenamefont {Gillot}, \citenamefont {Henry},
  \citenamefont {Hinderer}, \citenamefont {Holleville}, \citenamefont {Junca},
  \citenamefont {Lef{\`e}vre}, \citenamefont {Merzougui}, \citenamefont
  {Mielec}, \citenamefont {Monfret}, \citenamefont {Pelisson}, \citenamefont
  {Prevedelli}, \citenamefont {Reynaud}, \citenamefont {Riou}, \citenamefont
  {Rogister}, \citenamefont {Rosat}, \citenamefont {Cormier}, \citenamefont
  {Landragin}, \citenamefont {Chaibi}, \citenamefont {Gaffet},\ and\
  \citenamefont {Bouyer}}]{canuel_exploring_2018}%
  \BibitemOpen
  \bibfield  {author} {\bibinfo {author} {\bibfnamefont {B.}~\bibnamefont
  {Canuel}}, \bibinfo {author} {\bibfnamefont {A.}~\bibnamefont {Bertoldi}},
  \bibinfo {author} {\bibfnamefont {L.}~\bibnamefont {Amand}}, \bibinfo
  {author} {\bibfnamefont {E.}~\bibnamefont {Pozzo~di Borgo}}, \bibinfo
  {author} {\bibfnamefont {T.}~\bibnamefont {Chantrait}}, \bibinfo {author}
  {\bibfnamefont {C.}~\bibnamefont {Danquigny}}, \bibinfo {author}
  {\bibfnamefont {M.}~\bibnamefont {Dovale~{\'A}lvarez}}, \bibinfo {author}
  {\bibfnamefont {B.}~\bibnamefont {Fang}}, \bibinfo {author} {\bibfnamefont
  {A.}~\bibnamefont {Freise}}, \bibinfo {author} {\bibfnamefont
  {R.}~\bibnamefont {Geiger}}, \bibinfo {author} {\bibfnamefont
  {J.}~\bibnamefont {Gillot}}, \bibinfo {author} {\bibfnamefont
  {S.}~\bibnamefont {Henry}}, \bibinfo {author} {\bibfnamefont
  {J.}~\bibnamefont {Hinderer}}, \bibinfo {author} {\bibfnamefont
  {D.}~\bibnamefont {Holleville}}, \bibinfo {author} {\bibfnamefont
  {J.}~\bibnamefont {Junca}}, \bibinfo {author} {\bibfnamefont
  {G.}~\bibnamefont {Lef{\`e}vre}}, \bibinfo {author} {\bibfnamefont
  {M.}~\bibnamefont {Merzougui}}, \bibinfo {author} {\bibfnamefont
  {N.}~\bibnamefont {Mielec}}, \bibinfo {author} {\bibfnamefont
  {T.}~\bibnamefont {Monfret}}, \bibinfo {author} {\bibfnamefont
  {S.}~\bibnamefont {Pelisson}}, \bibinfo {author} {\bibfnamefont
  {M.}~\bibnamefont {Prevedelli}}, \bibinfo {author} {\bibfnamefont
  {S.}~\bibnamefont {Reynaud}}, \bibinfo {author} {\bibfnamefont
  {I.}~\bibnamefont {Riou}}, \bibinfo {author} {\bibfnamefont {Y.}~\bibnamefont
  {Rogister}}, \bibinfo {author} {\bibfnamefont {S.}~\bibnamefont {Rosat}},
  \bibinfo {author} {\bibfnamefont {E.}~\bibnamefont {Cormier}}, \bibinfo
  {author} {\bibfnamefont {A.}~\bibnamefont {Landragin}}, \bibinfo {author}
  {\bibfnamefont {W.}~\bibnamefont {Chaibi}}, \bibinfo {author} {\bibfnamefont
  {S.}~\bibnamefont {Gaffet}}, \ and\ \bibinfo {author} {\bibfnamefont
  {P.}~\bibnamefont {Bouyer}},\ }\href {\doibase 10.1038/s41598-018-32165-z}
  {\bibfield  {journal} {\bibinfo  {journal} {Scientific Reports}\ }\textbf
  {\bibinfo {volume} {8}},\ \bibinfo {pages} {14064} (\bibinfo {year}
  {2018})}\BibitemShut {NoStop}%
\bibitem [{\citenamefont {Karr}\ and\ \citenamefont
  {Marchand}(2019)}]{karr_progress_2019}%
  \BibitemOpen
  \bibfield  {author} {\bibinfo {author} {\bibfnamefont {J.-P.}\ \bibnamefont
  {Karr}}\ and\ \bibinfo {author} {\bibfnamefont {D.}~\bibnamefont
  {Marchand}},\ }\href {\doibase 10.1038/d41586-019-03364-z} {\bibfield
  {journal} {\bibinfo  {journal} {Nature}\ }\textbf {\bibinfo {volume} {575}},\
  \bibinfo {pages} {61} (\bibinfo {year} {2019})}\BibitemShut {NoStop}%
\bibitem [{\citenamefont {Norcia}\ \emph {et~al.}(2018)\citenamefont {Norcia},
  \citenamefont {Cline}, \citenamefont {Muniz}, \citenamefont {Robinson},
  \citenamefont {Hutson}, \citenamefont {Goban}, \citenamefont {Marti},
  \citenamefont {Ye},\ and\ \citenamefont {Thompson}}]{norcia_frequency_2018}%
  \BibitemOpen
  \bibfield  {author} {\bibinfo {author} {\bibfnamefont {M.~A.}\ \bibnamefont
  {Norcia}}, \bibinfo {author} {\bibfnamefont {J.~R.~K.}\ \bibnamefont
  {Cline}}, \bibinfo {author} {\bibfnamefont {J.~A.}\ \bibnamefont {Muniz}},
  \bibinfo {author} {\bibfnamefont {J.~M.}\ \bibnamefont {Robinson}}, \bibinfo
  {author} {\bibfnamefont {R.~B.}\ \bibnamefont {Hutson}}, \bibinfo {author}
  {\bibfnamefont {A.}~\bibnamefont {Goban}}, \bibinfo {author} {\bibfnamefont
  {G.~E.}\ \bibnamefont {Marti}}, \bibinfo {author} {\bibfnamefont
  {J.}~\bibnamefont {Ye}}, \ and\ \bibinfo {author} {\bibfnamefont {J.~K.}\
  \bibnamefont {Thompson}},\ }\href {\doibase 10.1103/PhysRevX.8.021036}
  {\bibfield  {journal} {\bibinfo  {journal} {Phys. Rev. X}\ }\textbf {\bibinfo
  {volume} {8}},\ \bibinfo {pages} {021036} (\bibinfo {year}
  {2018})}\BibitemShut {NoStop}%
\bibitem [{\citenamefont {Thorpe}\ \emph {et~al.}(2011)\citenamefont {Thorpe},
  \citenamefont {Rippe}, \citenamefont {Fortier}, \citenamefont {Kirchner},\
  and\ \citenamefont {Rosenband}}]{thorpe_frequency_2011}%
  \BibitemOpen
  \bibfield  {author} {\bibinfo {author} {\bibfnamefont {M.~J.}\ \bibnamefont
  {Thorpe}}, \bibinfo {author} {\bibfnamefont {L.}~\bibnamefont {Rippe}},
  \bibinfo {author} {\bibfnamefont {T.~M.}\ \bibnamefont {Fortier}}, \bibinfo
  {author} {\bibfnamefont {M.~S.}\ \bibnamefont {Kirchner}}, \ and\ \bibinfo
  {author} {\bibfnamefont {T.}~\bibnamefont {Rosenband}},\ }\href {\doibase
  10.1038/nphoton.2011.215} {\bibfield  {journal} {\bibinfo  {journal} {Nat.
  Photonics}\ }\textbf {\bibinfo {volume} {5}},\ \bibinfo {pages} {688}
  (\bibinfo {year} {2011})}\BibitemShut {NoStop}%
\bibitem [{\citenamefont {Cook}\ \emph {et~al.}(2015)\citenamefont {Cook},
  \citenamefont {Rosenband},\ and\ \citenamefont
  {Leibrandt}}]{cook_laser-frequency_2015}%
  \BibitemOpen
  \bibfield  {author} {\bibinfo {author} {\bibfnamefont {S.}~\bibnamefont
  {Cook}}, \bibinfo {author} {\bibfnamefont {T.}~\bibnamefont {Rosenband}}, \
  and\ \bibinfo {author} {\bibfnamefont {D.~R.}\ \bibnamefont {Leibrandt}},\
  }\href {\doibase 10.1103/PhysRevLett.114.253902} {\bibfield  {journal}
  {\bibinfo  {journal} {Phys. Rev. Lett.}\ }\textbf {\bibinfo {volume} {114}},\
  \bibinfo {pages} {253902} (\bibinfo {year} {2015})}\BibitemShut {NoStop}%
\bibitem [{\citenamefont {K{\'e}f{\'e}lian}\ \emph {et~al.}(2009)\citenamefont
  {K{\'e}f{\'e}lian}, \citenamefont {Jiang}, \citenamefont {Lemonde},\ and\
  \citenamefont {Santarelli}}]{kefelian_ultralow-frequency-noise_2009}%
  \BibitemOpen
  \bibfield  {author} {\bibinfo {author} {\bibfnamefont {F.}~\bibnamefont
  {K{\'e}f{\'e}lian}}, \bibinfo {author} {\bibfnamefont {H.}~\bibnamefont
  {Jiang}}, \bibinfo {author} {\bibfnamefont {P.}~\bibnamefont {Lemonde}}, \
  and\ \bibinfo {author} {\bibfnamefont {G.}~\bibnamefont {Santarelli}},\
  }\href {\doibase 10.1364/OL.34.000914} {\bibfield  {journal} {\bibinfo
  {journal} {Optics Letters}\ }\textbf {\bibinfo {volume} {34}},\ \bibinfo
  {pages} {914} (\bibinfo {year} {2009})}\BibitemShut {NoStop}%
\bibitem [{\citenamefont {Dong}\ \emph {et~al.}(2015)\citenamefont {Dong},
  \citenamefont {Hu}, \citenamefont {Huang}, \citenamefont {Ye}, \citenamefont
  {Qu}, \citenamefont {Li},\ and\ \citenamefont {Liu}}]{dong_subhertz_2015}%
  \BibitemOpen
  \bibfield  {author} {\bibinfo {author} {\bibfnamefont {J.}~\bibnamefont
  {Dong}}, \bibinfo {author} {\bibfnamefont {Y.}~\bibnamefont {Hu}}, \bibinfo
  {author} {\bibfnamefont {J.}~\bibnamefont {Huang}}, \bibinfo {author}
  {\bibfnamefont {M.}~\bibnamefont {Ye}}, \bibinfo {author} {\bibfnamefont
  {Q.}~\bibnamefont {Qu}}, \bibinfo {author} {\bibfnamefont {T.}~\bibnamefont
  {Li}}, \ and\ \bibinfo {author} {\bibfnamefont {L.}~\bibnamefont {Liu}},\
  }\href {\doibase 10.1364/AO.54.001152} {\bibfield  {journal} {\bibinfo
  {journal} {Appl. Opt.}\ }\textbf {\bibinfo {volume} {54}},\ \bibinfo {pages}
  {1152} (\bibinfo {year} {2015})}\BibitemShut {NoStop}%
\bibitem [{\citenamefont {Dong}\ \emph {et~al.}(2016)\citenamefont {Dong},
  \citenamefont {Huang}, \citenamefont {Li},\ and\ \citenamefont
  {Liu}}]{dong_observation_2016}%
  \BibitemOpen
  \bibfield  {author} {\bibinfo {author} {\bibfnamefont {J.}~\bibnamefont
  {Dong}}, \bibinfo {author} {\bibfnamefont {J.}~\bibnamefont {Huang}},
  \bibinfo {author} {\bibfnamefont {T.}~\bibnamefont {Li}}, \ and\ \bibinfo
  {author} {\bibfnamefont {L.}~\bibnamefont {Liu}},\ }\href {\doibase
  10.1063/1.4939918} {\bibfield  {journal} {\bibinfo  {journal} {Appl. Phys.
  Lett.}\ }\textbf {\bibinfo {volume} {108}},\ \bibinfo {pages} {021108}
  (\bibinfo {year} {2016})}\BibitemShut {NoStop}%
\bibitem [{\citenamefont {Lim}\ \emph {et~al.}(2017)\citenamefont {Lim},
  \citenamefont {Savchenkov}, \citenamefont {Dale}, \citenamefont {Liang},
  \citenamefont {Eliyahu}, \citenamefont {Ilchenko}, \citenamefont {Matsko},
  \citenamefont {Maleki},\ and\ \citenamefont {Wong}}]{lim_chasing_2017}%
  \BibitemOpen
  \bibfield  {author} {\bibinfo {author} {\bibfnamefont {J.}~\bibnamefont
  {Lim}}, \bibinfo {author} {\bibfnamefont {A.~A.}\ \bibnamefont {Savchenkov}},
  \bibinfo {author} {\bibfnamefont {E.}~\bibnamefont {Dale}}, \bibinfo {author}
  {\bibfnamefont {W.}~\bibnamefont {Liang}}, \bibinfo {author} {\bibfnamefont
  {D.}~\bibnamefont {Eliyahu}}, \bibinfo {author} {\bibfnamefont
  {V.}~\bibnamefont {Ilchenko}}, \bibinfo {author} {\bibfnamefont {A.~B.}\
  \bibnamefont {Matsko}}, \bibinfo {author} {\bibfnamefont {L.}~\bibnamefont
  {Maleki}}, \ and\ \bibinfo {author} {\bibfnamefont {C.~W.}\ \bibnamefont
  {Wong}},\ }\href {\doibase 10.1038/s41467-017-00021-9} {\bibfield  {journal}
  {\bibinfo  {journal} {Nature Communications}\ }\textbf {\bibinfo {volume}
  {8}},\ \bibinfo {pages} {8} (\bibinfo {year} {2017})}\BibitemShut {NoStop}%
\bibitem [{\citenamefont {Salomon}\ \emph {et~al.}(1988)\citenamefont
  {Salomon}, \citenamefont {Hils},\ and\ \citenamefont
  {Hall}}]{salomon_laser_1988}%
  \BibitemOpen
  \bibfield  {author} {\bibinfo {author} {\bibfnamefont {C.}~\bibnamefont
  {Salomon}}, \bibinfo {author} {\bibfnamefont {D.}~\bibnamefont {Hils}}, \
  and\ \bibinfo {author} {\bibfnamefont {J.~L.}\ \bibnamefont {Hall}},\ }\href
  {\doibase 10.1364/JOSAB.5.001576} {\bibfield  {journal} {\bibinfo  {journal}
  {J. Opt. Soc. Am. B}\ }\textbf {\bibinfo {volume} {5}},\ \bibinfo {pages}
  {1576} (\bibinfo {year} {1988})}\BibitemShut {NoStop}%
\bibitem [{\citenamefont {Webster}\ \emph {et~al.}(2008)\citenamefont
  {Webster}, \citenamefont {Oxborrow}, \citenamefont {Pugla}, \citenamefont
  {Millo},\ and\ \citenamefont {Gill}}]{webster_thermal-noise-limited_2008}%
  \BibitemOpen
  \bibfield  {author} {\bibinfo {author} {\bibfnamefont {S.~A.}\ \bibnamefont
  {Webster}}, \bibinfo {author} {\bibfnamefont {M.}~\bibnamefont {Oxborrow}},
  \bibinfo {author} {\bibfnamefont {S.}~\bibnamefont {Pugla}}, \bibinfo
  {author} {\bibfnamefont {J.}~\bibnamefont {Millo}}, \ and\ \bibinfo {author}
  {\bibfnamefont {P.}~\bibnamefont {Gill}},\ }\href {\doibase
  10.1103/PhysRevA.77.033847} {\bibfield  {journal} {\bibinfo  {journal} {Phys.
  Rev. A}\ }\textbf {\bibinfo {volume} {77}},\ \bibinfo {pages} {033847}
  (\bibinfo {year} {2008})}\BibitemShut {NoStop}%
\bibitem [{\citenamefont {Kessler}\ \emph {et~al.}(2012)\citenamefont
  {Kessler}, \citenamefont {Hagemann}, \citenamefont {Grebing}, \citenamefont
  {Legero}, \citenamefont {Sterr}, \citenamefont {Riehle}, \citenamefont
  {Martin}, \citenamefont {Chen},\ and\ \citenamefont
  {Ye}}]{kessler_sub-40-mhz-linewidth_2012}%
  \BibitemOpen
  \bibfield  {author} {\bibinfo {author} {\bibfnamefont {T.}~\bibnamefont
  {Kessler}}, \bibinfo {author} {\bibfnamefont {C.}~\bibnamefont {Hagemann}},
  \bibinfo {author} {\bibfnamefont {C.}~\bibnamefont {Grebing}}, \bibinfo
  {author} {\bibfnamefont {T.}~\bibnamefont {Legero}}, \bibinfo {author}
  {\bibfnamefont {U.}~\bibnamefont {Sterr}}, \bibinfo {author} {\bibfnamefont
  {F.}~\bibnamefont {Riehle}}, \bibinfo {author} {\bibfnamefont {M.~J.}\
  \bibnamefont {Martin}}, \bibinfo {author} {\bibfnamefont {L.}~\bibnamefont
  {Chen}}, \ and\ \bibinfo {author} {\bibfnamefont {J.}~\bibnamefont {Ye}},\
  }\href {\doibase 10.1038/nphoton.2012.217} {\bibfield  {journal} {\bibinfo
  {journal} {Nature Photonics}\ }\textbf {\bibinfo {volume} {6}},\ \bibinfo
  {pages} {687} (\bibinfo {year} {2012})}\BibitemShut {NoStop}%
\bibitem [{\citenamefont {Matei}\ \emph {et~al.}(2017)\citenamefont {Matei},
  \citenamefont {Legero}, \citenamefont {H{\"a}fner}, \citenamefont {Grebing},
  \citenamefont {Weyrich}, \citenamefont {Zhang}, \citenamefont {Sonderhouse},
  \citenamefont {Robinson}, \citenamefont {Ye}, \citenamefont {Riehle},\ and\
  \citenamefont {Sterr}}]{matei_1.5_2017}%
  \BibitemOpen
  \bibfield  {author} {\bibinfo {author} {\bibfnamefont {D.~G.}\ \bibnamefont
  {Matei}}, \bibinfo {author} {\bibfnamefont {T.}~\bibnamefont {Legero}},
  \bibinfo {author} {\bibfnamefont {S.}~\bibnamefont {H{\"a}fner}}, \bibinfo
  {author} {\bibfnamefont {C.}~\bibnamefont {Grebing}}, \bibinfo {author}
  {\bibfnamefont {R.}~\bibnamefont {Weyrich}}, \bibinfo {author} {\bibfnamefont
  {W.}~\bibnamefont {Zhang}}, \bibinfo {author} {\bibfnamefont
  {L.}~\bibnamefont {Sonderhouse}}, \bibinfo {author} {\bibfnamefont {J.~M.}\
  \bibnamefont {Robinson}}, \bibinfo {author} {\bibfnamefont {J.}~\bibnamefont
  {Ye}}, \bibinfo {author} {\bibfnamefont {F.}~\bibnamefont {Riehle}}, \ and\
  \bibinfo {author} {\bibfnamefont {U.}~\bibnamefont {Sterr}},\ }\href
  {\doibase 10.1103/PhysRevLett.118.263202} {\bibfield  {journal} {\bibinfo
  {journal} {Phys. Rev. Lett.}\ }\textbf {\bibinfo {volume} {118}},\ \bibinfo
  {pages} {263202} (\bibinfo {year} {2017})}\BibitemShut {NoStop}%
\bibitem [{\citenamefont {Numata}\ \emph {et~al.}(2004)\citenamefont {Numata},
  \citenamefont {Kemery},\ and\ \citenamefont
  {Camp}}]{numata_thermal-noise_2004}%
  \BibitemOpen
  \bibfield  {author} {\bibinfo {author} {\bibfnamefont {K.}~\bibnamefont
  {Numata}}, \bibinfo {author} {\bibfnamefont {A.}~\bibnamefont {Kemery}}, \
  and\ \bibinfo {author} {\bibfnamefont {J.}~\bibnamefont {Camp}},\ }\href
  {\doibase 10.1103/PhysRevLett.93.250602} {\bibfield  {journal} {\bibinfo
  {journal} {Phys. Rev. Lett.}\ }\textbf {\bibinfo {volume} {93}},\ \bibinfo
  {pages} {250602} (\bibinfo {year} {2004})}\BibitemShut {NoStop}%
\bibitem [{\citenamefont {Zhang}\ \emph {et~al.}(2017)\citenamefont {Zhang},
  \citenamefont {Robinson}, \citenamefont {Sonderhouse}, \citenamefont
  {Oelker}, \citenamefont {Benko}, \citenamefont {Hall}, \citenamefont
  {Legero}, \citenamefont {Matei}, \citenamefont {Riehle}, \citenamefont
  {Sterr},\ and\ \citenamefont {Ye}}]{zhang_ultrastable_2017}%
  \BibitemOpen
  \bibfield  {author} {\bibinfo {author} {\bibfnamefont {W.}~\bibnamefont
  {Zhang}}, \bibinfo {author} {\bibfnamefont {J.~M.}\ \bibnamefont {Robinson}},
  \bibinfo {author} {\bibfnamefont {L.}~\bibnamefont {Sonderhouse}}, \bibinfo
  {author} {\bibfnamefont {E.}~\bibnamefont {Oelker}}, \bibinfo {author}
  {\bibfnamefont {C.}~\bibnamefont {Benko}}, \bibinfo {author} {\bibfnamefont
  {J.~L.}\ \bibnamefont {Hall}}, \bibinfo {author} {\bibfnamefont
  {T.}~\bibnamefont {Legero}}, \bibinfo {author} {\bibfnamefont {D.~G.}\
  \bibnamefont {Matei}}, \bibinfo {author} {\bibfnamefont {F.}~\bibnamefont
  {Riehle}}, \bibinfo {author} {\bibfnamefont {U.}~\bibnamefont {Sterr}}, \
  and\ \bibinfo {author} {\bibfnamefont {J.}~\bibnamefont {Ye}},\ }\href
  {\doibase 10.1103/PhysRevLett.119.243601} {\bibfield  {journal} {\bibinfo
  {journal} {Phys. Rev. Lett.}\ }\textbf {\bibinfo {volume} {119}},\ \bibinfo
  {pages} {243601} (\bibinfo {year} {2017})}\BibitemShut {NoStop}%
\bibitem [{\citenamefont {Robinson}\ \emph {et~al.}(2019)\citenamefont
  {Robinson}, \citenamefont {Oelker}, \citenamefont {Milner}, \citenamefont
  {Zhang}, \citenamefont {Legero}, \citenamefont {Matei}, \citenamefont
  {Riehle}, \citenamefont {Sterr},\ and\ \citenamefont
  {Ye}}]{robinson_crystalline_2019}%
  \BibitemOpen
  \bibfield  {author} {\bibinfo {author} {\bibfnamefont {J.~M.}\ \bibnamefont
  {Robinson}}, \bibinfo {author} {\bibfnamefont {E.}~\bibnamefont {Oelker}},
  \bibinfo {author} {\bibfnamefont {W.~R.}\ \bibnamefont {Milner}}, \bibinfo
  {author} {\bibfnamefont {W.}~\bibnamefont {Zhang}}, \bibinfo {author}
  {\bibfnamefont {T.}~\bibnamefont {Legero}}, \bibinfo {author} {\bibfnamefont
  {D.~G.}\ \bibnamefont {Matei}}, \bibinfo {author} {\bibfnamefont
  {F.}~\bibnamefont {Riehle}}, \bibinfo {author} {\bibfnamefont
  {U.}~\bibnamefont {Sterr}}, \ and\ \bibinfo {author} {\bibfnamefont
  {J.}~\bibnamefont {Ye}},\ }\href {\doibase 10.1364/OPTICA.6.000240}
  {\bibfield  {journal} {\bibinfo  {journal} {Optica, OPTICA}\ }\textbf
  {\bibinfo {volume} {6}},\ \bibinfo {pages} {240} (\bibinfo {year}
  {2019})}\BibitemShut {NoStop}%
\bibitem [{\citenamefont {Matei}\ \emph {et~al.}(2016)\citenamefont {Matei},
  \citenamefont {Legero}, \citenamefont {Grebing}, \citenamefont {H{\"a}fner},
  \citenamefont {Lisdat}, \citenamefont {Weyrich}, \citenamefont {Zhang},
  \citenamefont {Sonderhouse}, \citenamefont {Robinson}, \citenamefont
  {Riehle}, \citenamefont {Ye},\ and\ \citenamefont
  {Sterr}}]{matei_second_2016}%
  \BibitemOpen
  \bibfield  {author} {\bibinfo {author} {\bibfnamefont {D.~G.}\ \bibnamefont
  {Matei}}, \bibinfo {author} {\bibfnamefont {T.}~\bibnamefont {Legero}},
  \bibinfo {author} {\bibfnamefont {C.}~\bibnamefont {Grebing}}, \bibinfo
  {author} {\bibfnamefont {S.}~\bibnamefont {H{\"a}fner}}, \bibinfo {author}
  {\bibfnamefont {C.}~\bibnamefont {Lisdat}}, \bibinfo {author} {\bibfnamefont
  {R.}~\bibnamefont {Weyrich}}, \bibinfo {author} {\bibfnamefont
  {W.}~\bibnamefont {Zhang}}, \bibinfo {author} {\bibfnamefont
  {L.}~\bibnamefont {Sonderhouse}}, \bibinfo {author} {\bibfnamefont {J.~M.}\
  \bibnamefont {Robinson}}, \bibinfo {author} {\bibfnamefont {F.}~\bibnamefont
  {Riehle}}, \bibinfo {author} {\bibfnamefont {J.}~\bibnamefont {Ye}}, \ and\
  \bibinfo {author} {\bibfnamefont {U.}~\bibnamefont {Sterr}},\ }\href
  {\doibase 10.1088/1742-6596/723/1/012031} {\bibfield  {journal} {\bibinfo
  {journal} {J. Phys.: Conf. Ser.}\ }\textbf {\bibinfo {volume} {723}},\
  \bibinfo {pages} {012031} (\bibinfo {year} {2016})}\BibitemShut {NoStop}%
\bibitem [{\citenamefont {White}(1975)}]{white_thermal_1975}%
  \BibitemOpen
  \bibfield  {author} {\bibinfo {author} {\bibfnamefont {G.~K.}\ \bibnamefont
  {White}},\ }\href {\doibase 10.1103/PhysRevLett.34.204} {\bibfield  {journal}
  {\bibinfo  {journal} {Phys. Rev. Lett.}\ }\textbf {\bibinfo {volume} {34}},\
  \bibinfo {pages} {204} (\bibinfo {year} {1975})}\BibitemShut {NoStop}%
\bibitem [{\citenamefont {Arcizet}\ \emph {et~al.}(2009)\citenamefont
  {Arcizet}, \citenamefont {Rivi{\`e}re}, \citenamefont {Schliesser},
  \citenamefont {Anetsberger},\ and\ \citenamefont
  {Kippenberg}}]{arcizet_cryogenic_2009}%
  \BibitemOpen
  \bibfield  {author} {\bibinfo {author} {\bibfnamefont {O.}~\bibnamefont
  {Arcizet}}, \bibinfo {author} {\bibfnamefont {R.}~\bibnamefont
  {Rivi{\`e}re}}, \bibinfo {author} {\bibfnamefont {A.}~\bibnamefont
  {Schliesser}}, \bibinfo {author} {\bibfnamefont {G.}~\bibnamefont
  {Anetsberger}}, \ and\ \bibinfo {author} {\bibfnamefont {T.~J.}\ \bibnamefont
  {Kippenberg}},\ }\href {\doibase 10.1103/PhysRevA.80.021803} {\bibfield
  {journal} {\bibinfo  {journal} {Phys. Rev. A}\ }\textbf {\bibinfo {volume}
  {80}},\ \bibinfo {pages} {021803} (\bibinfo {year} {2009})}\BibitemShut
  {NoStop}%
\bibitem [{\citenamefont {Zhu}\ \emph {et~al.}(2020)\citenamefont {Zhu},
  \citenamefont {Fokoua}, \citenamefont {Taranta}, \citenamefont {Chen},
  \citenamefont {Bradley}, \citenamefont {Petrovich}, \citenamefont {Poletti},
  \citenamefont {Zhao}, \citenamefont {Richardson},\ and\ \citenamefont
  {Slav{\'i}k}}]{zhu_thermal_2020}%
  \BibitemOpen
  \bibfield  {author} {\bibinfo {author} {\bibfnamefont {W.}~\bibnamefont
  {Zhu}}, \bibinfo {author} {\bibfnamefont {E.~R.~N.}\ \bibnamefont {Fokoua}},
  \bibinfo {author} {\bibfnamefont {A.~A.}\ \bibnamefont {Taranta}}, \bibinfo
  {author} {\bibfnamefont {Y.}~\bibnamefont {Chen}}, \bibinfo {author}
  {\bibfnamefont {T.}~\bibnamefont {Bradley}}, \bibinfo {author} {\bibfnamefont
  {M.~N.}\ \bibnamefont {Petrovich}}, \bibinfo {author} {\bibfnamefont
  {F.}~\bibnamefont {Poletti}}, \bibinfo {author} {\bibfnamefont
  {M.}~\bibnamefont {Zhao}}, \bibinfo {author} {\bibfnamefont {D.~J.}\
  \bibnamefont {Richardson}}, \ and\ \bibinfo {author} {\bibfnamefont
  {R.}~\bibnamefont {Slav{\'i}k}},\ }\href {\doibase 10.1109/JLT.2019.2960437}
  {\bibfield  {journal} {\bibinfo  {journal} {Journal of Lightwave Technology}\
  }\textbf {\bibinfo {volume} {38}},\ \bibinfo {pages} {2477} (\bibinfo {year}
  {2020})}\BibitemShut {NoStop}%
\bibitem [{\citenamefont {Black}(2001)}]{black_introduction_2001}%
  \BibitemOpen
  \bibfield  {author} {\bibinfo {author} {\bibfnamefont {E.~D.}\ \bibnamefont
  {Black}},\ }\href {\doibase 10.1119/1.1286663} {\bibfield  {journal}
  {\bibinfo  {journal} {American Journal of Physics}\ }\textbf {\bibinfo
  {volume} {69}},\ \bibinfo {pages} {79} (\bibinfo {year} {2001})}\BibitemShut
  {NoStop}%
\bibitem [{\citenamefont {Leibrandt}\ \emph {et~al.}(2011)\citenamefont
  {Leibrandt}, \citenamefont {Thorpe}, \citenamefont {Notcutt}, \citenamefont
  {Drullinger}, \citenamefont {Rosenband},\ and\ \citenamefont
  {Bergquist}}]{leibrandt_spherical_2011}%
  \BibitemOpen
  \bibfield  {author} {\bibinfo {author} {\bibfnamefont {D.~R.}\ \bibnamefont
  {Leibrandt}}, \bibinfo {author} {\bibfnamefont {M.~J.}\ \bibnamefont
  {Thorpe}}, \bibinfo {author} {\bibfnamefont {M.}~\bibnamefont {Notcutt}},
  \bibinfo {author} {\bibfnamefont {R.~E.}\ \bibnamefont {Drullinger}},
  \bibinfo {author} {\bibfnamefont {T.}~\bibnamefont {Rosenband}}, \ and\
  \bibinfo {author} {\bibfnamefont {J.~C.}\ \bibnamefont {Bergquist}},\ }\href
  {\doibase 10.1364/OE.19.003471} {\bibfield  {journal} {\bibinfo  {journal}
  {Opt. Express}\ }\textbf {\bibinfo {volume} {19}},\ \bibinfo {pages} {3471}
  (\bibinfo {year} {2011})}\BibitemShut {NoStop}%
\bibitem [{\citenamefont {Huang}\ \emph {et~al.}(2019)\citenamefont {Huang},
  \citenamefont {Wang}, \citenamefont {Duan}, \citenamefont {Huang},
  \citenamefont {Ye}, \citenamefont {Liu},\ and\ \citenamefont
  {Li}}]{huang_optical_2019}%
  \BibitemOpen
  \bibfield  {author} {\bibinfo {author} {\bibfnamefont {J.}~\bibnamefont
  {Huang}}, \bibinfo {author} {\bibfnamefont {L.}~\bibnamefont {Wang}},
  \bibinfo {author} {\bibfnamefont {Y.}~\bibnamefont {Duan}}, \bibinfo {author}
  {\bibfnamefont {Y.}~\bibnamefont {Huang}}, \bibinfo {author} {\bibfnamefont
  {M.}~\bibnamefont {Ye}}, \bibinfo {author} {\bibfnamefont {L.}~\bibnamefont
  {Liu}}, \ and\ \bibinfo {author} {\bibfnamefont {T.}~\bibnamefont {Li}},\
  }in\ \href {\doibase 10.1109/FCS.2019.8856066} {\emph {\bibinfo {booktitle}
  {2019 {Joint} {Conference} of {EFTF}/{IFC}}}}\ (\bibinfo {year} {2019})\ pp.\
  \bibinfo {pages} {1--3},\ \bibinfo {note} {iSSN: 2327-1949}\BibitemShut
  {NoStop}%
\bibitem [{\citenamefont {Egan}\ \emph {et~al.}(2015)\citenamefont {Egan},
  \citenamefont {Stone}, \citenamefont {Hendricks}, \citenamefont {Ricker},
  \citenamefont {Scace},\ and\ \citenamefont
  {Strouse}}]{egan_performance_2015}%
  \BibitemOpen
  \bibfield  {author} {\bibinfo {author} {\bibfnamefont {P.~F.}\ \bibnamefont
  {Egan}}, \bibinfo {author} {\bibfnamefont {J.~A.}\ \bibnamefont {Stone}},
  \bibinfo {author} {\bibfnamefont {J.~H.}\ \bibnamefont {Hendricks}}, \bibinfo
  {author} {\bibfnamefont {J.~E.}\ \bibnamefont {Ricker}}, \bibinfo {author}
  {\bibfnamefont {G.~E.}\ \bibnamefont {Scace}}, \ and\ \bibinfo {author}
  {\bibfnamefont {G.~F.}\ \bibnamefont {Strouse}},\ }\href {\doibase
  10.1364/OL.40.003945} {\bibfield  {journal} {\bibinfo  {journal} {Opt.
  Lett.}\ }\textbf {\bibinfo {volume} {40}},\ \bibinfo {pages} {3945} (\bibinfo
  {year} {2015})}\BibitemShut {NoStop}%
\bibitem [{\citenamefont {Wanser}(1992)}]{wanser_fundamental_1992}%
  \BibitemOpen
  \bibfield  {author} {\bibinfo {author} {\bibfnamefont {K.~H.}\ \bibnamefont
  {Wanser}},\ }\href {\doibase 10.1049/el:19920033} {\bibfield  {journal}
  {\bibinfo  {journal} {Electronics Letters}\ }\textbf {\bibinfo {volume}
  {28}},\ \bibinfo {pages} {53} (\bibinfo {year} {1992})}\BibitemShut {NoStop}%
\bibitem [{\citenamefont {Duan}(2010)}]{duan_intrinsic_2010}%
  \BibitemOpen
  \bibfield  {author} {\bibinfo {author} {\bibfnamefont {L.~Z.}\ \bibnamefont
  {Duan}},\ }\href {\doibase 10.1049/el.2010.1653} {\bibfield  {journal}
  {\bibinfo  {journal} {Electronics Letters}\ }\textbf {\bibinfo {volume}
  {46}},\ \bibinfo {pages} {1515} (\bibinfo {year} {2010})}\BibitemShut
  {NoStop}%
\bibitem [{\citenamefont {Merkel}\ \emph {et~al.}(2020)\citenamefont {Merkel},
  \citenamefont {Ulanowski},\ and\ \citenamefont
  {Reiserer}}]{merkel_coherent_2020}%
  \BibitemOpen
  \bibfield  {author} {\bibinfo {author} {\bibfnamefont {B.}~\bibnamefont
  {Merkel}}, \bibinfo {author} {\bibfnamefont {A.}~\bibnamefont {Ulanowski}}, \
  and\ \bibinfo {author} {\bibfnamefont {A.}~\bibnamefont {Reiserer}},\ }\href
  {\doibase 10.1103/PhysRevX.10.041025} {\bibfield  {journal} {\bibinfo
  {journal} {Phys. Rev. X}\ }\textbf {\bibinfo {volume} {10}},\ \bibinfo
  {pages} {041025} (\bibinfo {year} {2020})}\BibitemShut {NoStop}%
\bibitem [{\citenamefont {McRae}\ \emph {et~al.}(2013)\citenamefont {McRae},
  \citenamefont {Ngo}, \citenamefont {Shaddock}, \citenamefont {Hsu},\ and\
  \citenamefont {Gray}}]{mcrae_frequency_2013}%
  \BibitemOpen
  \bibfield  {author} {\bibinfo {author} {\bibfnamefont {T.~G.}\ \bibnamefont
  {McRae}}, \bibinfo {author} {\bibfnamefont {S.}~\bibnamefont {Ngo}}, \bibinfo
  {author} {\bibfnamefont {D.~A.}\ \bibnamefont {Shaddock}}, \bibinfo {author}
  {\bibfnamefont {M.~T.~L.}\ \bibnamefont {Hsu}}, \ and\ \bibinfo {author}
  {\bibfnamefont {M.~B.}\ \bibnamefont {Gray}},\ }\href {\doibase
  10.1364/OL.38.000278} {\bibfield  {journal} {\bibinfo  {journal} {Opt.
  Lett.}\ }\textbf {\bibinfo {volume} {38}},\ \bibinfo {pages} {278} (\bibinfo
  {year} {2013})}\BibitemShut {NoStop}%
\end{thebibliography}%

\end{document}